\documentclass[a4paper,11pt]{article}
\usepackage{pos}
\usepackage{lineno}
\usepackage{multirow} 
\usepackage{hyperref}

\title{New Structures in the $J/\psi$ $J/\psi$ Mass Spectrum at CMS}

\manuallySeparateAuthors
\author[a,b]{Xining Wang}
\author[]{ and } 
\author*[a,b]{Kai Yi}
\author{on behalf of the CMS Collaboration}

\affiliation[a]{Department of Physics, Tsinghua University,\\
  Haidian District, Beijing, China}
\affiliation[b]{School of Physics and Technology, Nanjing Normal University,\\
Wenyuan Road No. 1, Nanjing, China}

\emailAdd{xining.wang@cern.ch}
\emailAdd{yik@fnal.gov}

\abstract{A search is reported for structures near the $J/\psi$ $J/\psi$ mass threshold using a dataset of proton-proton collisions at $\sqrt{s} \: =13 \: \mathrm{TeV} $ recorded with the CMS detector at the LHC, corresponding to an integrated luminosity of about $135 \: \mathrm{fb^{-1}}$. Two structures are observed with a significance exceeding $5  \sigma$ and evidence of an additional structure is reported with a local significance of $4.7\sigma$.
}

\FullConference{22nd Conference on Flavor Physics and CP Violation (FPCP 2024)\\
 27-31 May 2024\\
 Bangkok, Thailand\\}

\begin{document}
\maketitle



\section{History of exotic hadrons}
Gell-Mann's original 1964 quark paper~\cite{Gell-Mann:1964ewy} introduced the possibility of exotic hadrons, which are states that differ from the usual $q\bar{q}$ or $qqq$ combinations. 
Since then, interest in observing these exotic states has fluctuated both theoretically and experimentally. 

In 2003, the Belle Collaboration made a significant discovery by identifying the $X(3872)$ state~\cite{Belle:2003nnu} (now referred to as $\chi_{c1}(3872)$). 
This discovery propelled exotic hadrons from speculative ideas to the forefront of research in hadron physics. 

Currently, while there are known candidates for doubly-heavy tetraquarks and heavy pentaquarks, 
their interpretations remain contentious. 
The main challenge lies in understanding the quark structure of these states, such as whether they can be modeled as molecules, diquarks, hybrids~\cite{Brambilla:2019esw},
or within a super-symmetric light front holographic QCD framework~\cite{Nielsen:2018ytt,Nielsen:2018uyn}. 
Some theorists even challenge the notion of bound-state interpretation 
and suggest that certain structures are artifacts of kinematic thresholds~\cite{ali_maiani_polosa_2019}.


\begin{figure}[h]
    \centering
    \includegraphics[width=0.48\textwidth]{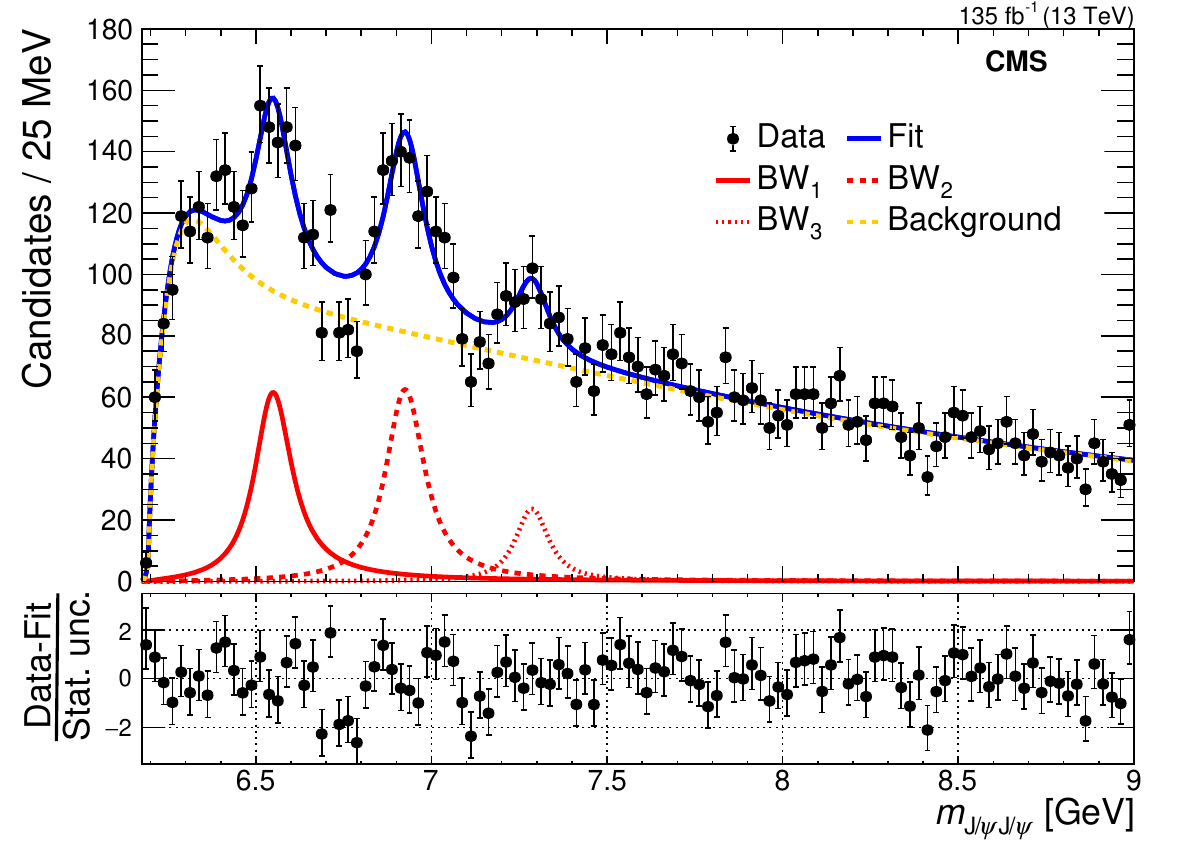}
    \includegraphics[width=0.48\textwidth]{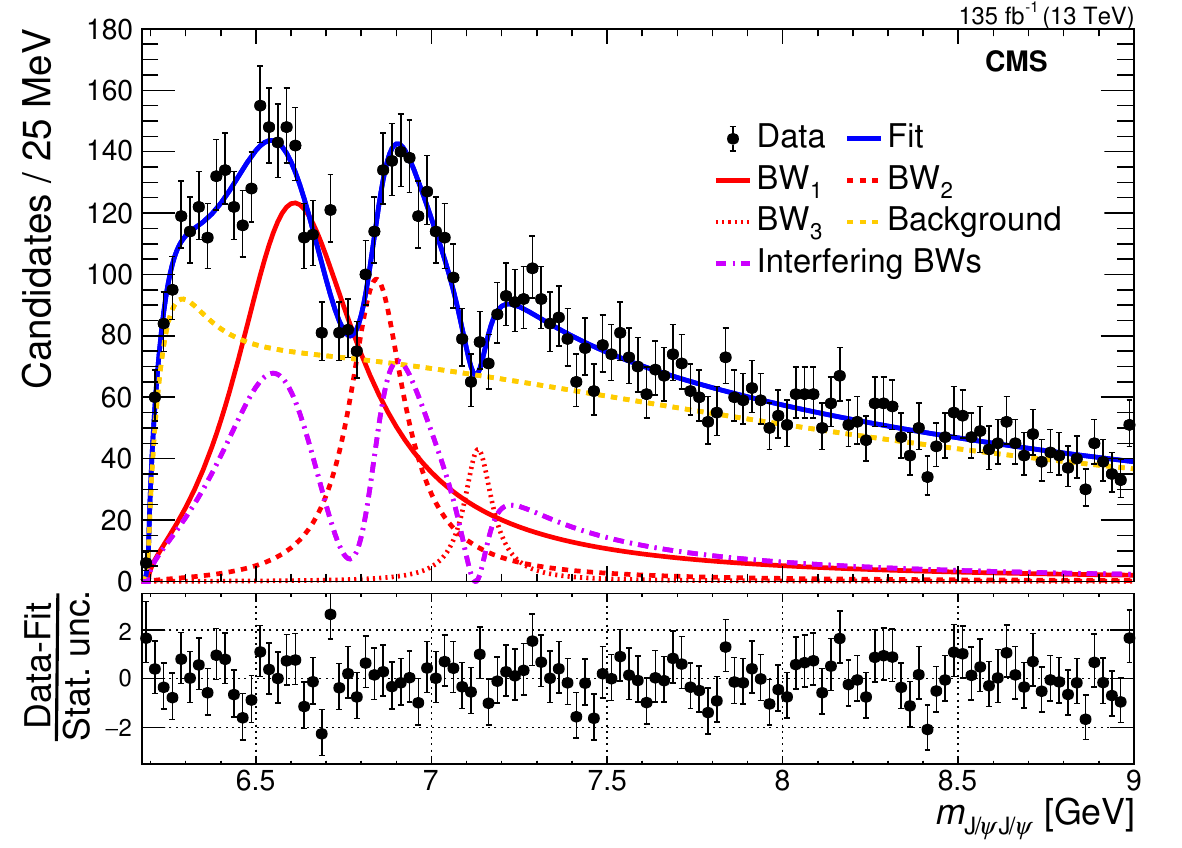}
    \caption{The fits to the $J/\psi J/\psi$ invariant mass spectrum in the CMS experiment~\cite{CMS:2023owd}. The fit model incorporates three signal functions ($X(6600)[\mathrm{BW_1}]$, $X(6900)[\mathrm{BW_2}]$, and $X(7300)[\mathrm{BW_3}]$) along with a background model. The fitting outcomes are presented in the left plot without considering interference effect, and in the right plot with interference included.}
    \label{fig:CMS}
\end{figure}

\section{All-charm tetraquark candidates}

In 2020, the LHCb Collaboration observed a structure in the $J/\psi J/\psi $ mass spectrum, 
named $X(6900)$~\cite{LHCb:2020bwg}.
It was subsequently confirmed by the ATLAS~\cite{ATLAS:2023bft} and CMS~\cite{CMS:2023owd} experiments, generating significant interest as a potential all-charm tetraquark~\cite{Wang:2020wrp,Berezhnoy:2011xn,Wu:2016vtq,Chen:2020xwe,Bai:2016int,Wang:2017jtz,Anwar:2017toa,Richard:2017vry,Esposito:2018cwh,Karliner:2016zzc,Bedolla:2019zwg,Anwar:2017toa,Yang:2020rih,Jin:2020jfc,Liu:2019zuc,liu:2020eha,Liu:2021rtn,Chen:2016jxd,Wang:2019rdo,Zhu:2020xni}. In addition, the CMS Collaboration reported the discovery of a new structure, $X(6600)$, with a local significance of 7.9 standard deviations, along with evidence for another new structure provisionally named $X(7100)$, exhibiting a local significance of 4.7 standard deviations.

The search in the CMS experiment focuses on the $J/\psi J/\psi$ mass spectrum from proton-proton collisions, with $J/\psi$ reconstructed from $\mu^{+}\mu^{-}$~\cite{CMS:2023owd}. 
The observed $J/\psi J/\psi$ spectrum and the fits, both without and with consideration of interference effects, are illustrated in Fig.~\ref{fig:CMS} and discussed in detail below.

The CMS detector is well-suited for studying exotic quarkonium states, thanks to its high-purity muon identification, excellent mass resolution for $J/\psi$, and precise vertex resolution. Moreover, specialized triggers based on muons are utilized to select quarkonium candidates. A more detailed description of the CMS detector can be found in Ref.~\cite{CMS:2008xjf}.




\section{Fit without interference}

In order to extract resonance parameters, a fitting package has been designed to conduct the fit. The amplitude of the signal resonance is modeled by the relativistic Breit-Wigner function~\cite{Zyla:2020zbs, Bohm:2004zi}:
\begin{align}
  & BW(m;m_0,\Gamma_0) = \frac{\sqrt{m\Gamma(m)}}{m_0^2 - m^2 - i m\Gamma(m)}, \\
  & \Gamma(m) = \Gamma_0 \left(\frac{q}{q_0}\right)^{2L + 1}\frac{m_0}{m}\left(B^{\prime}_L(q, q_0, d)\right)^2 , \\
  & B^\prime_L(q, q_0, d)
  = \frac{q^{-L}B_L(q,d)}{q_0^{-L}B_L(q_0,d)}
  = \left(\frac{q_0}{q}\right)^L\frac{B_L(q,d)}{B_L(q_0,d)}, \\
  & B_0(q, d) = 1, \\
  & B_1(q, d) = \sqrt{\frac{2z}{z+1}},\\
  & B_2(q, d) = \sqrt{\frac{13z^2}{(z-3)^2 + 9z}},\\
  & z = (|q|d)^2, z_0 = (|q_0|d)^2,
\end{align}
where $q$ represents the magnitude of momentum of a daughter particle in the resonance rest frame, while $L$ denotes the orbital angular momentum number between the two daughters, and the subscript $0$ indicates the value at the peak mass. The default fit utilizes $L=0$, but we explored other values of $L$ as a part of systematic uncertainty studies.
The term $B_L(q, d)$ corresponds to the Blatt-Weisskopf barrier factor~\cite{Hippel:1972, PhysRevD.48.1225}. The parameter $d$ is set to be 3 GeV$^{-1}$ ($\sim$ 0.6 fm) as employed in Ref.~\cite{Aaij:2015tga}. We varied the value of $d$ in the studies of the systematic uncertainty.

Simple summations of Breit-Wigner functions are employed to fit the $J/\psi J/\psi$ mass spectrum in both LHCb and CMS experiments, with the line-shape parameters summarized in Table~\ref{tab:nointerf}. The fitting results are depicted in
Fig.~\ref{fig:CMS} (left) for CMS. 
However, these fits inadequately capture the $J/\psi J/\psi$ mass spectrum, as they fail to accurately describe the dips between the peaks.

\begin{table}[]
    \centering
    \caption{The line-shape parameters of the states extracted from no-interference fits 
    in the LHCb and CMS experiments. The first uncertainties are statistical while the second ones are systematic.}
    \label{tab:nointerf}
    \begin{tabular}{lllll}
    \hline
    \multicolumn{2}{l}{}  & $X(6600)$  & $X(6900)$  & $X(7100)$ \\
    \hline
    \multirow{2}{*}{LHCb} & $m\,[\mathrm{MeV}]$  &    & $6905 \pm 11 \pm 7$       &    \\
                          & $\Gamma\,[\mathrm{MeV}]$ & & $80 \pm 19 \pm 33$      &    \\
    \hline
    \multirow{2}{*}{CMS}  & $m\,[\mathrm{MeV}]$  &  $6552 \pm 10 \pm 12$    & $6927 \pm 9 \pm 4$    &  $7287^{+20}_{-18} \pm 5$  \\
                          & $\Gamma\,[\mathrm{MeV}]$ & $124 ^{+32}_{-26} \pm 33$ & $122 ^{+24}_{-21} \pm 18$ &  $95 ^{+59}_{-40} \pm 19$ \\
    \hline
    \end{tabular}
\end{table}

\section{Fit with interference}

It's important to note that the dips in the $J/\psi J/\psi $ mass spectrum are observed in all LHCb, CMS and ATLAS experiments, which are poorly described by the no-interfering Breit-Wigner functions in LHCb and CMS experiments.
Various ways to employ interference among fitting components were considered by the different experiments in order to explain the dips --- but other explanations are conceivable.

In the CMS experiment, two dips are observed around $6750\,\mathrm{MeV}$ and $7150\,\mathrm{MeV}$, and three structures are found in the $J/\psi J/\psi $ mass spectrum~\cite{CMS:2020qwa}. Interference models are developed to elucidate these dips.
After investigating potential interference among the different components, the primary interference fit model is determined to involve the interference between three resonances $X(6600)$[BW1], $X(6900)$[BW2], and $X(7100)$[BW3], implemented with a term proportional to $\left| r_{1} \exp{(i\phi_{1})}\mathrm{BW_{1}} + \mathrm{BW_{2}} + r_{3}\exp{(i\phi_{3})}\mathrm{BW_{3}} \right|$, where $r_{1,3}$ and $\phi_{1,3}$ denote the relative magnitudes and phases of $\mathrm{BW_{1,3}}$ with respect to $\mathrm{BW_{2}}$. 
The fit outcome is illustrated in Fig.~\ref{fig:CMS} (right), demonstrating an improvement in the $\chi^{2}$ probability in the signal region [6.2, 7.8] $\mathrm{GeV}$ to 65\%, compared to only 9\% in the no-interference fit.
While the interference between pairs of components is also considered, their signal-region $\chi^{2}$ probabilities remain below 30\%, leading to their exclusion as the nominal interference fit model.

To account for the dip, the LHCb experiment incorporated interference 
between the non-resonant single parton scattering (NRSPS) background 
with the resonant production of an auxillary BW~\cite{LHCb:2020bwg}.
Analogous to LHCb approach, the ATLAS experiment constructed a model, 
referred as Model B, with a lower broad structure interfering with the NRSPS background.
Another ATLAS model, denoted as Model A, 
introduced interference among the $X(6900)$ and two lower mass resonances.

The line-shape parameters of the states extracted from the interference fits
in the three experiments are summarized in Table~\ref{tab:interf}. Notably, the measured mass and width of $X(6900)$ are found to be comparable across all three experiments.
The interference fit indicates that the unidentified $J^{PC}$ quantum numbers of these states could be identical, suggesting they may originate from a coherent production process.

\begin{table}[]
    \centering
    \caption{The line-shape parameters of the states extracted from interference fit 
    in LHCb, CMS and ATLAS experiments. Two kinds of interference models are considered in ATLAS experiment, 
    named as Model A and Model B. The first uncertainties are statistical while the second ones are systematic.}
    \label{tab:interf}
    \begin{tabular}{llllll}
    \hline
    \multicolumn{2}{l}{}    &   &   $X(6600)$ & $X(6900)$ & $X(7100)$ \\
    \hline
    \multicolumn{2}{l}{\multirow{2}{*}{LHCb}} & $m\,[\mathrm{MeV}]$      &   & $6886 \pm 11 \pm 11$     &    \\
    \multicolumn{2}{l}{}                       & $\Gamma\,[\mathrm{MeV}]$ & & $168 \pm 33 \pm 69$  &       \\
    \hline
    \multicolumn{2}{l}{\multirow{2}{*}{CMS}}   & $m\,[\mathrm{MeV}]$   &  $6638 ^{+43+16}_{-38-31}$    & $6847 ^{+44+48}_{-28-20}$    &   $7134^{+48+41}_{-25-15}$  \\
    \multicolumn{2}{l}{}                       & $\Gamma\,[\mathrm{MeV}]$ & $440 ^{+230+110}_{-200-240}$ & $191 ^{+66+25}_{-49-17}$  &  $97 ^{+40+29}_{-29-26} $ \\
    \hline
    \multirow{4}{*}{ATLAS} & \multirow{2}{*}{Model A} & $m\,[\mathrm{MeV}]$  &  &   $6860 \pm 30 ^{+10}_{-20}$    &    \\
                           &                          & $\Gamma\,[\mathrm{MeV}]$ &   &   $110 \pm 50 ^{+20}_{-10}$    &    \\
                           & \multirow{2}{*}{Model B} & $m\,[\mathrm{MeV}]$  &   &    $6910 \pm 10 \pm 10$   &       \\
                           &                          & $\Gamma\,[\mathrm{MeV}]$ &   &   $150 \pm 30 \pm 10$  &      \\
    \hline
    \end{tabular}
\end{table}

\section{Comparison with some theoretical calculations}
The measured masses in the CMS experiment from both no-interference and interference fits 
appear compatible with recent calculations of the $cc\bar{c}\bar{c}$ spectrum~\cite{zhu:Wu_2024,Zhu:2020xni,Tiwari:2021tmz}, the correctness of which will be determined by the preference of nature for no-interference or interference case.
These three structures may be a family of radial excitations of the same $J^{PC}$,
which is the case for both no-interference and interference masses, albeit for different theoretical models.

Many theoretical models predict the quantum numbers for $X(6900)$,
including as a spin-0 state~\cite{Zhu:2020xni}, or a spin-2 state~\cite{Ma:2020kwb}.
The measurement of the spin and parity of these states is deemed crucial in distinguishing between competing theoretical models, enabling theorists to perform calculations based on accurate assumptions regarding the quantum numbers, finding out their position in the tetracharm spectroscopy, and better understanding the nature of exotic hadrons.

\section{Summary}
In summary, the analysis of the $J/\psi J/\psi$ invariant mass spectrum from proton-proton collisions at $\sqrt{s}=13\, \mathrm{TeV}$ using the CMS detector, based on an integrated luminosity of $135\, fb^{-1}$, has revealed three distinct structures. These structures are effectively described by a model incorporating interference between three resonances. Among these findings, two new structures, provisionally designated as $X(6600)$ and $X(7100)$, have been identified with local statistical significances of 7.9 and 4.7 standard deviations, respectively. The observation of $X(6900)$ confirms the discovery by the LHCb experiment, which exhibited a local significance of 9.8 standard deviations measured by the CMS experiment.

\bibliographystyle{JHEP}
\bibliography{myrefs}

\providecommand{\href}[2]{#2}\begingroup\raggedright\begin{thebibliography}{10}

\bibitem{Gell-Mann:1964ewy}
M.~Gell-Mann, \emph{A schematic model of baryons and mesons},
  \href{https://doi.org/10.1016/S0031-9163(64)92001-3}{\emph{Phys. Lett.}
  {\bfseries 8} (1964) 214}.

\bibitem{Belle:2003nnu}
{\scshape Belle} collaboration, S.~K. Choi et~al., \emph{{Observation of a
  narrow charmonium-like state in exclusive $B^\pm \to K^\pm \pi^+ \pi^-
  J/\psi$ decays}},
  \href{https://doi.org/10.1103/PhysRevLett.91.262001}{\emph{Phys. Rev. Lett.}
  {\bfseries 91} (2003) 262001}
  [\href{https://arxiv.org/abs/hep-ex/0309032}{{\ttfamily hep-ex/0309032}}].

\bibitem{Brambilla:2019esw}
N.~Brambilla, S.~Eidelman, C.~Hanhart, A.~Nefediev, C.-P. Shen, C.~E. Thomas
  et~al., \emph{{The $XYZ$ states: experimental and theoretical status and
  perspectives}},
  \href{https://doi.org/10.1016/j.physrep.2020.05.001}{\emph{Phys. Rept.}
  {\bfseries 873} (2020) 1} [\href{https://arxiv.org/abs/1907.07583}{{\ttfamily
  1907.07583}}].

\bibitem{Nielsen:2018ytt}
M.~Nielsen, S.~J. Brodsky, G.~F. de~T\'eramond, H.~G. Dosch, F.~S. Navarra and
  L.~Zou, \emph{{Supersymmetry in the Double-Heavy Hadronic Spectrum}},
  \href{https://doi.org/10.1103/PhysRevD.98.034002}{\emph{Phys. Rev. D}
  {\bfseries 98} (2018) 034002}
  [\href{https://arxiv.org/abs/1805.11567}{{\ttfamily 1805.11567}}].

\bibitem{Nielsen:2018uyn}
M.~Nielsen and S.~J. Brodsky, \emph{{Hadronic superpartners from a
  superconformal and supersymmetric algebra}},
  \href{https://doi.org/10.1103/PhysRevD.97.114001}{\emph{Phys. Rev. D}
  {\bfseries 97} (2018) 114001}
  [\href{https://arxiv.org/abs/1802.09652}{{\ttfamily 1802.09652}}].

\bibitem{ali_maiani_polosa_2019}
A.~Ali, L.~Maiani and A.~D. Polosa, \emph{Multiquark Hadrons}. Cambridge
  University Press, Cambridge, 2019,
  \href{https://doi.org/10.1017/9781316761465}{10.1017/9781316761465}.

\bibitem{CMS:2023owd}
{\scshape CMS} collaboration, A.~Hayrapetyan et~al., \emph{{New Structures in
  the J/\ensuremath{\psi}J/\ensuremath{\psi} Mass Spectrum in Proton-Proton
  Collisions at s=13\,\,TeV}},
  \href{https://doi.org/10.1103/PhysRevLett.132.111901}{\emph{Phys. Rev. Lett.}
  {\bfseries 132} (2024) 111901}
  [\href{https://arxiv.org/abs/2306.07164}{{\ttfamily 2306.07164}}].

\bibitem{LHCb:2020bwg}
{\scshape LHCb} collaboration, R.~Aaij et~al., \emph{{Observation of structure
  in the $J /\psi$ -pair mass spectrum}},
  \href{https://doi.org/10.1016/j.scib.2020.08.032}{\emph{Sci. Bull.}
  {\bfseries 65} (2020) 1983}
  [\href{https://arxiv.org/abs/2006.16957}{{\ttfamily 2006.16957}}].

\bibitem{ATLAS:2023bft}
{\scshape ATLAS} collaboration, G.~Aad et~al., \emph{{Observation of an Excess
  of Dicharmonium Events in the Four-Muon Final State with the ATLAS
  Detector}}, \href{https://doi.org/10.1103/PhysRevLett.131.151902}{\emph{Phys.
  Rev. Lett.} {\bfseries 131} (2023) 151902}
  [\href{https://arxiv.org/abs/2304.08962}{{\ttfamily 2304.08962}}].

\bibitem{Wang:2020wrp}
J.-Z. Wang, D.-Y. Chen, X.~Liu and T.~Matsuki, \emph{{Producing fully charm
  structures in the $J/\psi$ -pair invariant mass spectrum}},
  \href{https://doi.org/10.1103/PhysRevD.103.L071503}{\emph{Phys. Rev. D}
  {\bfseries 103} (2021) 071503}
  [\href{https://arxiv.org/abs/2008.07430}{{\ttfamily 2008.07430}}].

\bibitem{Berezhnoy:2011xn}
A.~V. Berezhnoy, A.~V. Luchinsky and A.~A. Novoselov, \emph{Heavy tetraquarks
  production at the {LHC}},
  \href{https://doi.org/10.1103/PhysRevD.86.034004}{\emph{Phys. Rev. D}
  {\bfseries 86} (2012) 034004}
  [\href{https://arxiv.org/abs/1111.1867}{{\ttfamily 1111.1867}}].

\bibitem{Wu:2016vtq}
J.~Wu, Y.-R. Liu, K.~Chen, X.~Liu and S.-L. Zhu, \emph{{Heavy-flavored
  tetraquark states with the $QQ\bar{Q}\bar{Q}$ configuration}},
  \href{https://doi.org/10.1103/PhysRevD.97.094015}{\emph{Phys. Rev. D}
  {\bfseries 97} (2018) 094015}
  [\href{https://arxiv.org/abs/1605.01134}{{\ttfamily 1605.01134}}].

\bibitem{Chen:2020xwe}
H.-X. Chen, W.~Chen, X.~Liu and S.-L. Zhu, \emph{Strong decays of fully-charm
  tetraquarks into di-charmonia},
  \href{https://doi.org/10.1016/j.scib.2020.08.038}{\emph{Sci. Bull.}
  {\bfseries 65} (2020) 1994}
  [\href{https://arxiv.org/abs/2006.16027}{{\ttfamily 2006.16027}}].

\bibitem{Bai:2016int}
Y.~Bai, S.~Lu and J.~Osborne, \emph{Beauty-full tetraquarks},
  \href{https://doi.org/10.1016/j.physletb.2019.134930}{\emph{Phys. Lett. B}
  {\bfseries 798} (2019) 134930}
  [\href{https://arxiv.org/abs/1612.00012}{{\ttfamily 1612.00012}}].

\bibitem{Wang:2017jtz}
Z.-G. Wang, \emph{{Analysis of the $QQ\bar{Q}\bar{Q}$ tetraquark states with
  QCD sum rules}},
  \href{https://doi.org/10.1140/epjc/s10052-017-4997-0}{\emph{Eur. Phys. J. C}
  {\bfseries 77} (2017) 432}
  [\href{https://arxiv.org/abs/1701.04285}{{\ttfamily 1701.04285}}].

\bibitem{Anwar:2017toa}
M.~N. Anwar, J.~Ferretti, F.-K. Guo, E.~Santopinto and B.-S. Zou,
  \emph{Spectroscopy and decays of the fully-heavy tetraquarks},
  \href{https://doi.org/10.1140/epjc/s10052-018-6073-9}{\emph{Eur. Phys. J. C}
  {\bfseries 78} (2018) 647}
  [\href{https://arxiv.org/abs/1710.02540}{{\ttfamily 1710.02540}}].

\bibitem{Richard:2017vry}
J.-M. Richard, A.~Valcarce and J.~Vijande, \emph{String dynamics and
  metastability of all-heavy tetraquarks},
  \href{https://doi.org/10.1103/PhysRevD.95.054019}{\emph{Phys. Rev. D}
  {\bfseries 95} (2017) 054019}
  [\href{https://arxiv.org/abs/1703.00783}{{\ttfamily 1703.00783}}].

\bibitem{Esposito:2018cwh}
A.~Esposito and A.~D. Polosa, \emph{{A $bb\bar b\bar b$ di-bottomonium at the
  LHC?}}, \href{https://doi.org/10.1140/epjc/s10052-018-6269-z}{\emph{Eur.
  Phys. J. C} {\bfseries 78} (2018) 782}
  [\href{https://arxiv.org/abs/1807.06040}{{\ttfamily 1807.06040}}].

\bibitem{Karliner:2016zzc}
M.~Karliner, S.~Nussinov and J.~L. Rosner, \emph{{$Q Q \bar Q \bar Q$ states:
  masses, production, and decays}},
  \href{https://doi.org/10.1103/PhysRevD.95.034011}{\emph{Phys. Rev. D}
  {\bfseries 95} (2017) 034011}
  [\href{https://arxiv.org/abs/1611.00348}{{\ttfamily 1611.00348}}].

\bibitem{Bedolla:2019zwg}
M.~A. Bedolla, J.~Ferretti, C.~D. Roberts and E.~Santopinto, \emph{Spectrum of
  fully-heavy tetraquarks from a diquark+antidiquark perspective},
  \href{https://doi.org/10.1140/epjc/s10052-020-08579-3}{\emph{Eur. Phys. J. C}
  {\bfseries 80} (2020) 1004}
  [\href{https://arxiv.org/abs/1911.00960}{{\ttfamily 1911.00960}}].

\bibitem{Yang:2020rih}
G.~Yang, J.~Ping, L.~He and Q.~Wang, \emph{{Potential model prediction of
  fully-heavy tetraquarks $QQ\bar{Q}\bar{Q}$ ($Q=c, b$)}},
  \href{https://arxiv.org/abs/2006.13756}{{\ttfamily 2006.13756}}.

\bibitem{Jin:2020jfc}
X.~Jin, Y.~Xue, H.~Huang and J.~Ping, \emph{{Full-heavy tetraquarks in
  constituent quark models}},
  \href{https://doi.org/10.1140/epjc/s10052-020-08650-z}{\emph{Eur. Phys. J. C}
  {\bfseries 80} (2020) 1083}
  [\href{https://arxiv.org/abs/2006.13745}{{\ttfamily 2006.13745}}].

\bibitem{Liu:2019zuc}
M.-S. Liu, Q.-F. L\"u, X.-H. Zhong and Q.~Zhao, \emph{{All-heavy tetraquarks}},
  \href{https://doi.org/10.1103/PhysRevD.100.016006}{\emph{Phys. Rev. D}
  {\bfseries 100} (2019) 016006}
  [\href{https://arxiv.org/abs/1901.02564}{{\ttfamily 1901.02564}}].

\bibitem{liu:2020eha}
M.-S. liu, F.-X. Liu, X.-H. Zhong and Q.~Zhao, \emph{{Full-heavy tetraquark
  states and their evidences in the LHCb di-$J/\psi$ spectrum}},
  \href{https://arxiv.org/abs/2006.11952}{{\ttfamily 2006.11952}}.

\bibitem{Liu:2021rtn}
F.-X. Liu, M.-S. Liu, X.-H. Zhong and Q.~Zhao, \emph{{Higher mass spectra of
  the fully-charmed and fully-bottom tetraquarks}},
  \href{https://doi.org/10.1103/PhysRevD.104.116029}{\emph{Phys. Rev. D}
  {\bfseries 104} (2021) 116029}
  [\href{https://arxiv.org/abs/2110.09052}{{\ttfamily 2110.09052}}].

\bibitem{Chen:2016jxd}
W.~Chen, H.-X. Chen, X.~Liu, T.~G. Steele and S.-L. Zhu, \emph{{Hunting for
  exotic doubly hidden-charm/bottom tetraquark states}},
  \href{https://doi.org/10.1016/j.physletb.2017.08.034}{\emph{Phys. Lett. B}
  {\bfseries 773} (2017) 247}
  [\href{https://arxiv.org/abs/1605.01647}{{\ttfamily 1605.01647}}].

\bibitem{Wang:2019rdo}
G.-J. Wang, L.~Meng and S.-L. Zhu, \emph{{Spectrum of the fully-heavy
  tetraquark state $QQ\bar Q' \bar Q'$}},
  \href{https://doi.org/10.1103/PhysRevD.100.096013}{\emph{Phys. Rev. D}
  {\bfseries 100} (2019) 096013}
  [\href{https://arxiv.org/abs/1907.05177}{{\ttfamily 1907.05177}}].

\bibitem{Zhu:2020xni}
R.~Zhu, \emph{{Fully-heavy tetraquark spectra and production at hadron
  colliders}},
  \href{https://doi.org/10.1016/j.nuclphysb.2021.115393}{\emph{Nucl. Phys. B}
  {\bfseries 966} (2021) 115393}
  [\href{https://arxiv.org/abs/2010.09082}{{\ttfamily 2010.09082}}].

\bibitem{CMS:2008xjf}
{\scshape CMS} collaboration, S.~Chatrchyan et~al., \emph{{The CMS Experiment
  at the CERN LHC}},
  \href{https://doi.org/10.1088/1748-0221/3/08/S08004}{\emph{JINST} {\bfseries
  3} (2008) S08004}.

\bibitem{Zyla:2020zbs}
{\scshape Particle Data Group} collaboration, P.~Zyla et~al., \emph{{Review of
  Particle Physics}}, \href{https://doi.org/10.1093/ptep/ptaa104}{\emph{PTEP}
  {\bfseries 2020} (2020) 083C01}.

\bibitem{Bohm:2004zi}
A.~R. Bohm and Y.~Sato, \emph{{Relativistic resonances: Their masses, widths,
  lifetimes, superposition, and causal evolution}},
  \href{https://doi.org/10.1103/PhysRevD.71.085018}{\emph{Phys. Rev. D}
  {\bfseries 71} (2005) 085018}
  [\href{https://arxiv.org/abs/hep-ph/0412106}{{\ttfamily hep-ph/0412106}}].

\bibitem{Hippel:1972}
F.~Von~Hippel and C.~Quigg, \emph{{Centrifugal-barrier effects in resonance
  partial decay widths, shapes, and production amplitudes}},
  \href{https://doi.org/10.1103/PhysRevD.5.624}{\emph{Phys. Rev. D} {\bfseries
  5} (1972) 624}.

\bibitem{PhysRevD.48.1225}
S.~U. Chung, \emph{{Helicity coupling amplitudes in tensor formalism}},
  \href{https://doi.org/10.1103/PhysRevD.56.4419}{\emph{Phys. Rev. D}
  {\bfseries 48} (1993) 1225}.

\bibitem{Aaij:2015tga}
{\scshape LHCb} collaboration, R.~Aaij et~al., \emph{{Observation of $J/\psi p$
  Resonances Consistent with Pentaquark States in $\Lambda_b^0 \to J/\psi K^-
  p$ Decays}},
  \href{https://doi.org/10.1103/PhysRevLett.115.072001}{\emph{Phys. Rev. Lett.}
  {\bfseries 115} (2015) 072001}
  [\href{https://arxiv.org/abs/1507.03414}{{\ttfamily 1507.03414}}].

\bibitem{CMS:2020qwa}
{\scshape CMS} collaboration, A.~M. Sirunyan et~al., \emph{{Measurement of the
  $\Upsilon$(1S) pair production cross section and search for resonances
  decaying to $\Upsilon$(1S)$\mu^+\mu^-$ in proton-proton collisions at
  $\sqrt{s} =$ 13 TeV}},
  \href{https://doi.org/10.1016/j.physletb.2020.135578}{\emph{Phys. Lett. B}
  {\bfseries 808} (2020) 135578}
  [\href{https://arxiv.org/abs/2002.06393}{{\ttfamily 2002.06393}}].

\bibitem{zhu:Wu_2024}
W.-L. Wu, Y.-K. Chen, L.~Meng and S.-L. Zhu, \emph{Benchmark calculations of
  fully heavy compact and molecular tetraquark states},
  \href{https://doi.org/10.1103/physrevd.109.054034}{\emph{Physical Review D}
  {\bfseries 109} (2024) }.

\bibitem{Tiwari:2021tmz}
R.~Tiwari, D.~P. Rathaud and A.~K. Rai, \emph{{Spectroscopy of all charm
  tetraquark states}},
  \href{https://doi.org/10.1007/s12648-022-02427-8}{\emph{Indian J. Phys.}
  {\bfseries 97} (2023) 943}
  [\href{https://arxiv.org/abs/2108.04017}{{\ttfamily 2108.04017}}].

\bibitem{Ma:2020kwb}
Y.-Q. Ma and H.-F. Zhang, \emph{{Exploring the Di-$J/\psi$ Resonances around
  6.9 $\mathrm{GeV}$ Based on $ab$ $initio$ Perturbative QCD}},
  \href{https://arxiv.org/abs/2009.08376}{{\ttfamily 2009.08376}}.

\end{thebibliography}\endgroup

\end{document}